\documentclass[useAMS,usenatbib]{mn2e}
\usepackage{graphicx,color,epsfig}

\newcommand{\be}{\begin{equation}}
\newcommand{\ee}{\end{equation}}

\newcommand{\apj}{ApJ}
\newcommand{\apjs}{ApJS}
\newcommand{\mnras}{MNRAS}
\newcommand{\aap}{A\&A}
\newcommand{\araa}{ARA\&A}
\newcommand{\apjl}{ApJL}
\newcommand{\aj}{AJ}

\def\ltsima{$\; \buildrel < \over \sim \;$}
\def\simlt{\lower.5ex\hbox{\ltsima}}
\def\gtsima{$\; \buildrel > \over \sim \;$}
\def\simgt{\lower.5ex\hbox{\gtsima}}

\def\msun{{\,{\rm M}_\odot}}

\def\del#1{{}}
\title[Competitive feedback in galaxy formation]
{Competitive feedback in galaxy formation}

\author[S. Nayakshin, M.I Wilkinson and A.R. King]{Sergei Nayakshin, Mark
  I. Wilkinson, and Andrew King \\ Department of Physics \& Astronomy,
  University of Leicester, Leicester, LE1 7RH, UK}
\begin{document}

\date{Received}

\pagerange{\pageref{firstpage}--\pageref{lastpage}} \pubyear{2008}

\maketitle

\label{firstpage}

\begin{abstract}

It is now well established that many galaxies have nuclear star clusters (NCs)
whose total masses correlate with the velocity dispersion $\sigma$ of the
galaxy spheroid in a very similar way to the well--known supermassive black
hole (SMBH) $M - \sigma$ relation. Previous theoretical work suggested that
both correlations can be explained by a momentum feedback argument.
Observations further show that most known NCs have masses $\la 10^8\msun$,
while SMBHs frequently have measured masses $\ga 10^8\msun$, which remained
unexplained in earlier treatments.  We suggest here that this changeover
reflects a competition between the SMBH and nuclear clusters in the feedback
they produce. When one of the massive objects reaches its limiting $M-\sigma$
value, it drives the gas away and hence cuts off its own mass and also the
mass of the ``competitor''. The latter is then underweight with respect to the
expected $M-\sigma$ mass.

More specifically, we find that the bulge dynamical timescale is a steeply
rising function of velocity dispersion, and that the NC--SMBH changeover occurs
where the dynamical time is about equal to the Salpeter time. We propose that
SMBHs, growing on the Salpeter time scale, are unable to reach their $M -
\sigma$ mass quickly enough in small bulges. The central regions of these
bulges are swamped with gas which fragments into stars, creating the nuclear
clusters. The latter then limit their own growth by the feedback they produce,
settling on their (offset) $M - \sigma$ relation. The SMBH in such bulges
should be underweight as their growth is curtailed before they reach the $M -
\sigma$ mass. In large bulges, on the other hand, the SMBH catches up quickly
enough to settle on its $M-\sigma$ relation. Nuclear star clusters may also
exist in such bulges but they should be underweight with respect to their
$M-\sigma$ sequence.
\end{abstract}


\section{Introduction}\label{intro}

It is well known that the masses of the supermassive black holes (SMBHs) in
the nuclei of early--type galaxies and bulges correlate with the velocity
dispersions of the stellar spheroids \citep[e.g.,][]{Gebhardt00, Ferrarese00,
  Tremaine02}. A simple explanation invokes momentum feedback
\citep{King03,King05}. In this picture the SMBH luminosity is limited by the
Eddington value, and the momentum outflow rate produced by radiation pressure
is of the order of
\begin{equation}
\dot\Pi_{\rm SMBH} \approx {L_{\rm Edd}\over c} = \frac{ 4 \pi G M_{\rm
    BH}}{\kappa}\;,
\label{pismbh}
\end{equation}
where $\kappa$ is the opacity, assumed to be dominated by the electron
scattering, and $M_{\rm BH}$ is the SMBH mass. This momentum flux
produces an outward force on the gas in the bulge, whose weight is
$W(R) = GM(R)[M_{\rm total}(R)]/ R^2$, where $M(R)$ is the enclosed
gas mass at radius $R$, and $M_{\rm total}(R)$ is the total enclosed
mass including dark matter. For an isothermal potential, $M(R)$ and
$M_{\rm total}(R)$ are proportional to $R$, so the result is
\begin{equation}
W = {4f_g\sigma^4\over G}\;.
\label{w}
\end{equation}
Here $f_g$ is the baryonic fraction and $\sigma^2 = GM/2R$ is
the velocity dispersion in the bulge. To order of magnitude, the
relation \ref{w} holds for any potential if estimated at the virial
radius.

Requiring that momentum output produced by the black hole should just
balance the weight of the gas leads to the $M_{\rm BH}$--$\sigma$
relation \citep{King03,King05}:
\begin{equation}
M_{\rm BH} = {f_g\kappa\over \pi G^2}\sigma^4.
\label{msig}
\end{equation}
The model is attractive in its physical simplicity. Further, the
result contains no free parameters, but is very close to the observed
$M_{\rm BH}$--$\sigma$ relation.

Another feature commonly found in the centres of galaxies are nuclear star
clusters. They are found in late type spirals \citep[e.g.,][]{BokerEtal02},
bulgeless spirals \citep{WalcherEtal05}, edge-on spirals \citep{SethEtal06},
and dwarf elliptical galaxies \citep{CoteEtal06}.  The cluster masses range
from $10^6$ to $10^8\msun$ \citep[although even more massive clusters have
  been found recently by][]{KormendyEtal09}, whereas their sizes are only a
few to a few tens parsec. Intriguingly, for dwarf ellipticals the masses of
NCs correlate with the host properties. Namely, \citet{FerrareseEtal06} found
that the NC mass is related to the bulge velocity dispersion in exactly same
way as the SMBH -- velocity dispersion relation, but with normalisation offset
by about an order of magnitude. The cluster masses are also almost linearly
proportional to the total bulge mass \citep{FerrareseEtal06, Wehner06}.

\cite{McLaughlinEtal06} proposed that the observed $M_{\rm NC} -\sigma$
relation for dwarf elliptical galaxies follows naturally from an extension of
the above argument (\cite{King03,King05}) to the outflows from young star
clusters containing massive stars. These individual stars are also
Eddington--limited, and produce outflows with momentum outflow rate $\sim
L_{\rm Edd}/c$ where $L_{\rm Edd}$ is calculated from the star's mass.  Young
star clusters with normal IMFs produce momentum outflow rate
\begin{equation}
\dot\Pi_{\rm NC} \approx \lambda{L_{\rm Edd}\over c}
\label{pinc}
\end{equation}
where $\lambda \approx 0.05$ and $L_{\rm Edd}$ is now formally the
Eddington value corresponding to the total cluster mass. To produce
the same amount of momentum feedback, a young star cluster must
therefore be $1/\lambda$ times more massive than a SMBH radiating at
the Eddington limit, and hence:
\begin{equation}
M_{\rm NC} = {f_g\kappa\over \lambda\pi G^2}\sigma^4.
\label{msignc}
\end{equation}
Strikingly, $1/\lambda$ is quite close to the offset in mass between
the $M_{\rm BH}$--$\sigma$ and $M_{\rm NC}$--$\sigma$
relations. Furthermore, while equation \ref{pismbh} is very plausible
\citep[see][]{KP03}, the momentum outflow rates from stars are known
in detail observationally. The largest uncertainty in the equation
\ref{pinc} is therefore the stellar IMF, which is observationally
fairly constant \citep[e.g.,][]{Kroupa02}. The \cite{McLaughlinEtal06}
explanation of $M_{\rm NC}$--$\sigma$ relation thus appears similarly
robust to the \cite{King03,King05} model for SMBH feedback.

However, \cite{McLaughlinEtal06} did not offer an explanation of why bulges
with smaller $\sigma$ contain nuclear clusters, while more massive galaxies
contain SMBHs and not NCs. Here we propose an explanation, noting that
timescales are important in this problem as well as energetics.

Our simple theory for the observed bimodality of NC and SMBHs is based on the
premise that the dominant object must be able to grow quickly and yet stay
active for long enough to provide the needed feedback. As we show below, in
small bulges this argument favours nuclear star clusters whereas in larger
ones the situation is reversed. Below we explain our idea, address
observational constraints and conditions needed for it to work, and suggest
possible astrophysical implications.

We note that nuclear star clusters could in principle form elsewhere in the
galaxies and then migrate inwards due to dynamical friction with the
background stars. However, \cite{Milosavljevic04a} argues against this
possibility due to the short time scales available for this process, and
argues instead that these clusters may form in situ. We agree with this point,
and further note that observations of young massive stars in the central
parsec of the Milky Way offer direct support to the in-situ formation model
\citep{NC05,PaumardEtal06,NS05}. The exact geometrical arrangement of the
forming stars (a thick disc or a quasi-spherical cluster) is irrelevant on the
scales of the parent galaxy.

\section{Timescales}\label{sec:argument}

Black holes and nuclear clusters each evolve on characteristic
timescales. SMBH growth is limited by the Eddington accretion rate,
$\dot M_{\rm Edd} = L_{\rm Edd}/(\epsilon c^2)$, where $\epsilon \sim
0.1$ is the radiative efficiency of accretion.  SMBH masses can grow
no faster than $\exp(t/t_{\rm Salp})$, where
\begin{equation}
t_{\rm Salp} = \frac{M_{\rm BH}}{\dot M_{\rm
  Edd}} = \frac{\kappa \epsilon c}{4\pi G} = 4.5\times
10^7\epsilon_{0.1}~{\rm yr}
\label{tsalp}
\end{equation}
is the Salpeter time, with $\epsilon_{0.1} = \epsilon/0.1$. Star
formation can occur on the free--fall or dynamical timescale $t_{\rm
  dyn}$ of the system, which is less than a million years for many
observed young star clusters \citep[e.g.,][]{Hillenbrand97}.

Once a SMBH is created, its feedback can be activated at any time, provided
that the accretion rate is high enough. By contrast, star cluster feedback has
a ``half life'' of around $t_{\rm MS} \simlt 2\times 10^7$~yr, since this is
the main--sequence lifetime of the massive stars contributing most to the
feedback \citep[e.g.][]{Leitherer1992}.  This timescale is only a factor of
two shorter than the Salpeter time. After this time ($t=t_{\rm ms}$), the
ability of the nuclear clusters to expel gas from the galaxy is severely
reduced. They would have to be rebuild their population of massive young stars
to restart. It is not obvious that this is physically possible inside an
existing dense stellar cluster.

Consider a bulge where the dynamical time (equation \ref{tvir} below)
is much shorter than the Salpeter time. In a gas feeding event (e.g. a
merger), the bulge regains dynamical equilibrium before any
significant SMBH growth and feedback sets in, and there is nothing to
prevent gas from collecting in the bulge centre. The accumulated gas
is then consumed by star formation in nuclear regions, forming nuclear
clusters which quickly reach their $M_{\rm NC}$--$\sigma$ limiting
mass. This cuts off growth of everything -- the bulge, the NC and the
SMBH as well. The SMBH in these bulges are thus bound to be
underweight compared to the $M_{\rm BH}$--$\sigma$ relation.

In the opposite extreme, when the bulge dynamical time is longer than
the Salpeter time, the SMBH can grow quickly enough to reach its
limiting $M_{\rm BH}$--$\sigma$ mass. While nuclear star clusters
might be created there as well, their feedback quickly (i.e. in about
20 million years) becomes negligible. The situation is thus the
reverse of the last paragraph, and it is the nuclear star clusters
that are underweight in these bulges.

Below we estimate $t_{\rm dyn}$ as a function of bulge mass or velocity
dispersion. We find that $t_{\rm Salp} \ga t_{\rm dyn}$ in smaller bulges
($\sigma \la 150$~km\,s$^{-1}$) and $t_{\rm Salp} \la t_{\rm dyn}$ in larger
ones.

\section{Dynamical time and  velocity dispersion}
\label{sec:tms}

A tacit but obvious assumption in the arguments of
\cite{King03,King05} and \cite{McLaughlinEtal06} is that the source of
feedback can respond quickly to bulge growth and thus influence it.
Depletion of gas in the bulge by star formation and the onset of
stellar feedback probably occur within a few bulge dynamical time
scales
\begin{equation}
t_{\rm dyn} = \frac{R}{\sigma}
\label{tvir}
\end{equation}
where $R$ and $\sigma$ are the scale length and velocity dispersion of
the bulge. If the feedback source fails to reach its limiting mass its
feedback remains unimportant.

We now consider how the dynamical time $t_{\rm dyn}$ scales with the
$\sigma$ of the stellar component of a galaxy. It is well established
that stellar spheroids occupy a two-dimensional `fundamental plane' in
the space defined by the total luminosity, the scale length and the
velocity dispersion~\citep{Djorgovski1987,Bernardi2003}. The plane is
tilted relative to its position expected from the virial theorem for a
homologous population of galaxies in a dynamical equilibrium. Several
explanations for the origin of the tilt have been discussed in the
literature, including systematic variations in the mass to light ratio
of the stellar populations, or changes in the dark matter
fraction~\citep[e.g.][]{Dekel2006}. Recent studies have shown that the
tilt is essentially independent of the wavelength of the observations,
suggesting that stellar population variations are not the dominant
contribution~\citep{LaBarbera2008,Bernardi2003}.

Projections of the fundamental plane lead to a number of simple
scaling relations. Using Sloan Digital Sky Survey photometric and
spectroscopic data for $\sim9000$ galaxies with measured velocity
dispersions of 100-400 km\,s$^{-1}$, \cite{Bernardi2003} derived the
following relations between $R$, $\sigma$ and the total luminosity
$L$: 
\begin{eqnarray}
R &=& 2.6 \left(\frac{L}{1.6\times10^{10}L_\odot}\right)^{0.704 \pm
  0.025}\, {\rm kpc}\\ \sigma &=& 150
\left(\frac{L}{1.6\times10^{10}L_\odot}\right)^{0.23 \pm 0.012}\, {\rm
  km\,s}^{-1}\label{eq:sigma_L}
\end{eqnarray}
(The power-law indices are taken from~\cite{Dekel2006} who report updated
values obtained by Bernardi (priv. comm.) using revised SDSS photometry.)
These relations lead to another one between the dynamical time and the total
luminosity, given by
\begin{equation}
t_{\rm dyn} = \frac{R}{\sigma} = 17
\left(\frac{L}{1.6\times10^{10}L_\odot}\right)^{0.474} \mbox{Myr}
\label{eq:tdyn}
\end{equation}
Combining relations~(\ref{eq:sigma_L}) and~(\ref{eq:tdyn}), we obtain the
dynamical time as a function of velocity dispersion, namely
\begin{equation}
t_{\rm dyn} =
17\left(\frac{\sigma}{150\,\mbox{km\,s}^{-1}}\right)^{2.06} \mbox{Myr}
\end{equation}
We have taken $\sigma \sim 150\,\mbox{km\,s}^{-1}$ as our fiducial value as
the observations show that no nuclear clusters have been observed in systems
with $\sigma \ga 150\,\mbox{km\,s}^{-1}$. The above relations show that this
roughly coincides with the transition between systems with dynamical times
longer than the Salpeter time~(\ref{tsalp}). The details of the transition may
depend on the merger history of the galaxy.

\section{Discussion}\label{sec:discussion}

We have seen that momentum feedback gives a simple physical
explanation of why galaxy bulges are dominated by nuclear clusters for
low velocity dispersions and by supermassive black holes for high
dispersions.  We have emphasised that given an injection of gas,
e.g. from a merger, galaxies with dynamical times shorter than the
Salpeter time cannot grow their central black holes sufficiently
quickly to affect the gas infall. Gas accumulating in the central
regions cannot cool and condense indefinitely, so nuclear star
clusters form and produce feedback. The masses of these clusters
saturate at the mass (\ref{msignc}) when they expel the remaining
gas. The hole thus remains close to its `seed' mass, which is
presumably less than the value (\ref{msig}).

Note that this line of argument does not imply that dwarf elliptical galaxies
with low velocity dispersion do not contain massive black holes. We only
suggest here that their growth is slow; it is not entirely
forbidden. Therefore, these galaxies may still contain underweight SMBH, i.e.,
black holes with mass significantly less than the corresponding $M_{\rm
  BH}$--$\sigma$ value. It seems difficult to avoid building up a massive
black hole in the very centre of the cluster and galaxy potential
well. Mergers of low mass holes may provide an interesting additional window
for gravitational wave astronomy
\citep{2004ApJ...614..864M,2009ApJ...692L..50A}.

There is an important constraint for our model to be applicable. Star
formation in the inner parts of galaxies with higher velocity dispersions is
not ruled out by the considerations of this paper. These galaxies could thus
potentially build up nuclear clusters in their centres. If this happens faster
than the Salpeter time, then the masses of these clusters should saturate at
the value (\ref{msig}). Presumably, if the SMBH continues to grow at the
Eddington rate, it could then reach its limiting $M_{\rm BH}$--$\sigma$
mass. However this picture would predict very massive nuclear star clusters
(up to $M_{\rm NC} \sim 10^{10} \msun$ for $\sigma \sim 300$ km/sec) which are
not observed.

This suggests that SMBH growth by accretion should be the dominant process in
the central parsecs of galaxies, and that star formation occurs only as an
alternative when gas cannot be consumed by the hole quickly enough. The latter
naturally occurs if (a) the material is first deposited into the disc on small
scales where star formation does not occur due to the strong SMBH tidal effect
\citep[e.g.,][]{Kolykhalov80,KingPringle07}, and (b) the hole is fed at a
super-Eddington rate. Then the hole accretes the gas at the Eddington rate,
expelling the rest. A good fraction of the expelled gas would probably not
travel very far from the centre of the galaxy. Gas is likely to be expelled
with a range of velocities, some too low to escape to infinity. As its angular
momentum is very low, the gas can fall back into the inner parsec(s). Such
effects are actually observed in the simulations of accretion discs winds by
\cite{Proga03c}.  Deposited back into the accretion disc on parsec scales, the
gas would then fuel star formation there \citep[e.g.,][]{Goodman03}.

Summarising this, applicability of our model demands that SMBH feeding be
primary and star formation secondary in the inner few parsecs of AGN. If
this holds, nuclear star clusters grow only when SMBHs cannot.

Finally, our model explicitly assumes that the nuclear star clusters and
bulges of dwarf ellipticals form in a quasi-spherical or at least a
geometrically thick disc configuration of gas. If instead the gas is in a
thin disc configuration before the onset of star formation, the feedback
efficiency would be greatly reduced, and no significant bulge would be formed.
Therefore our model does not apply to bulgeless spiral galaxies. If
central $\simlt $ tens of parsec of these galaxies are fed via gaseous discs
\citep[e.g.,][]{Milosavljevic04a}, then the mass of the NCs need not saturate
at the value given by equation \ref{msignc}.

The assumption that star formation proceeds on a single dynamical timescale is
a lower limit on the time actually required.  In fact it is more likely that
star formation in the bulge takes several dynamical times to complete. Indeed,
observationally we know that Giant Molecular Clouds in the Milky Way must be
contracting much slower than dynamical collapse \citep{ZuckermanPalmer74} to
explain the low star formation efficiency in the Galaxy, presumably due to
feedback by star formation inside the clouds \citep{McKee89}. We would
therefore expect a transition regime around $\sigma \sim
150\,\mbox{km\,s}^{-1}$ where galaxies may contain either nuclear clusters or
black holes. This boundary region extends over a factor of $\sim 2$ in
$\sigma$. In this region the competition between NCs and SMBH depends on the
detail of gas deposition in the inner region of the galaxy and perhaps the
merger history of the galaxy.

The picture we have presented is necessarily very simplified. One
would like to include effects such as a realistic galaxy bulge
potential, density inhomogeneities, and possible cooling
effects. Further, the changeover between NC and SMBH--dominated bulges
depends on the merger history of the galaxy. For all these reasons a
numerical treatment of this picture is desirable. The fact that our
own Galaxy appears to lie in the regime where the merger history may
play an important role should make such studies rewarding.

\section{Acknowledgments}

Theoretical astrophysics research at the University of Leicester is
supported by a STFC Rolling grant. MIW acknowledges a Royal Society
University Research Fellowship.

%\bibliographystyle{mnras}
%\bibliography{../nayakshin}

\begin{thebibliography}{33}
\expandafter\ifx\csname natexlab\endcsname\relax\def\natexlab#1{#1}\fi

\bibitem[{Amaro-Seoane} et~al.(2009){Amaro-Seoane}, {Miller} \&
  {Freitag}]{2009ApJ...692L..50A}
{Amaro-Seoane} P., {Miller} M.~C., {Freitag} M., 2009, \apjl, 692, L50

\bibitem[{Bernardi} et~al.(2003){Bernardi}, {Sheth}, {Annis}
  et~al.]{Bernardi2003}
{Bernardi} M., {Sheth} R.~K., {Annis} J., et~al., 2003, \aj, 125, 1849

\bibitem[{B{\"o}ker} et~al.(2002){B{\"o}ker}, {Laine}, {van der Marel}
  et~al.]{BokerEtal02}
{B{\"o}ker} T., {Laine} S., {van der Marel} R.~P., et~al., 2002, \aj, 123, 1389

\bibitem[{C{\^o}t{\'e}} et~al.(2006){C{\^o}t{\'e}}, {Piatek}, {Ferrarese}
  et~al.]{CoteEtal06}
{C{\^o}t{\'e}} P., {Piatek} S., {Ferrarese} L., et~al., 2006, \apjs, 165, 57

\bibitem[{Dekel} \& {Cox}(2006)]{Dekel2006}
{Dekel} A., {Cox} T.~J., 2006, \mnras, 370, 1445

\bibitem[{Djorgovski} \& {Davis}(1987)]{Djorgovski1987}
{Djorgovski} S., {Davis} M., 1987, \apj, 313, 59

\bibitem[{Ferrarese} et~al.(2006){Ferrarese}, {C{\^o}t{\'e}}, {Dalla Bont{\`a}}
  et~al.]{FerrareseEtal06}
{Ferrarese} L., {C{\^o}t{\'e}} P., {Dalla Bont{\`a}} E., et~al., 2006, \apjl,
  644, L21

\bibitem[{Ferrarese} \& {Merritt}(2000)]{Ferrarese00}
{Ferrarese} L., {Merritt} D., 2000, \apjl, 539, L9

\bibitem[{Gebhardt} et~al.(2000){Gebhardt}, {Bender}, {Bower}
  et~al.]{Gebhardt00}
{Gebhardt} K., {Bender} R., {Bower} G., et~al., 2000, \apjl, 539, L13

\bibitem[{Goodman}(2003)]{Goodman03}
{Goodman} J., 2003, \mnras, 339, 937

\bibitem[{Hillenbrand}(1997)]{Hillenbrand97}
{Hillenbrand} L.~A., 1997, \aj, 113, 1733

\bibitem[{King}(2003)]{King03}
{King} A., 2003, \apjl, 596, L27

\bibitem[{King}(2005)]{King05}
{King} A., 2005, \apjl, 635, L121

\bibitem[{King} \& {Pounds}(2003)]{KP03}
{King} A.~R., {Pounds} K.~A., 2003, \mnras, 345, 657

\bibitem[{King} \& {Pringle}(2007)]{KingPringle07}
{King} A.~R., {Pringle} J.~E., 2007, \mnras, 377, L25

\bibitem[{Kolykhalov} \& {Sunyaev}(1980)]{Kolykhalov80}
{Kolykhalov} P.~I., {Sunyaev} R.~A., 1980, Soviet Astron. Lett., 6, 357

\bibitem[{Kormendy} et~al.(2009){Kormendy}, {Fisher}, {Cornell} \&
  {Bender}]{KormendyEtal09}
{Kormendy} J., {Fisher} D.~B., {Cornell} M.~E., {Bender} R., 2009, \apjs, 182,
  216

\bibitem[{Kroupa}(2002)]{Kroupa02}
{Kroupa} P., 2002, Science, 295, 82

\bibitem[{La Barbera} et~al.(2008){La Barbera}, {Busarello}, {Merluzzi}, {de la
  Rosa}, {Coppola} \& {Haines}]{LaBarbera2008}
{La Barbera} F., {Busarello} G., {Merluzzi} P., {de la Rosa} I.~G., {Coppola}
  G., {Haines} C.~P., 2008, \apj, 689, 913

\bibitem[{Leitherer} et~al.(1992){Leitherer}, {Robert} \&
  {Drissen}]{Leitherer1992}
{Leitherer} C., {Robert} C., {Drissen} L., 1992, \apj, 401, 596

\bibitem[{Matsubayashi} et~al.(2004){Matsubayashi}, {Shinkai} \&
  {Ebisuzaki}]{2004ApJ...614..864M}
{Matsubayashi} T., {Shinkai} H.-a., {Ebisuzaki} T., 2004, \apj, 614, 864

\bibitem[{McKee}(1989)]{McKee89}
{McKee} C.~F., 1989, \apj, 345, 782

\bibitem[{McLaughlin} et~al.(2006){McLaughlin}, {King} \&
  {Nayakshin}]{McLaughlinEtal06}
{McLaughlin} D.~E., {King} A.~R., {Nayakshin} S., 2006, \apjl, 650, L37

\bibitem[{Milosavljevi{\'c}}(2004)]{Milosavljevic04a}
{Milosavljevi{\'c}} M., 2004, \apjl, 605, L13

\bibitem[{Nayakshin} \& {Cuadra}(2005)]{NC05}
{Nayakshin} S., {Cuadra} J., 2005, \aap, 437, 437

\bibitem[{Nayakshin} \& {Sunyaev}(2005)]{NS05}
{Nayakshin} S., {Sunyaev} R., 2005, \mnras, 364, L23

\bibitem[{Paumard} et~al.(2006){Paumard}, {Genzel}, {Martins}
  et~al.]{PaumardEtal06}
{Paumard} T., {Genzel} R., {Martins} F., et~al., 2006, \apj, 643, 1011

\bibitem[{Proga}(2003)]{Proga03c}
{Proga} D., 2003, \apj, 585, 406

\bibitem[{Seth} et~al.(2006){Seth}, {Dalcanton}, {Hodge} \&
  {Debattista}]{SethEtal06}
{Seth} A.~C., {Dalcanton} J.~J., {Hodge} P.~W., {Debattista} V.~P., 2006, \aj,
  132, 2539

\bibitem[{Tremaine} et~al.(2002){Tremaine}, {Gebhardt}, {Bender}
  et~al.]{Tremaine02}
{Tremaine} S., {Gebhardt} K., {Bender} R., et~al., 2002, \apj, 574, 740

\bibitem[{Walcher} et~al.(2005){Walcher}, {van der Marel}, {McLaughlin}
  et~al.]{WalcherEtal05}
{Walcher} C.~J., {van der Marel} R.~P., {McLaughlin} D., et~al., 2005, \apj,
  618, 237

\bibitem[{Wehner} \& {Harris}(2006)]{Wehner06}
{Wehner} E.~H., {Harris} W.~E., 2006, \apjl, 644, L17

\bibitem[{Zuckerman} \& {Palmer}(1974)]{ZuckermanPalmer74}
{Zuckerman} B., {Palmer} P., 1974, \araa, 12, 279

\end{thebibliography}

\label{lastpage}

\end{document}